\documentstyle[12pt]{article}
\begin{document}

\begin{titlepage}
\begin{flushright}
Alberta Thy-27-94\\hep-ph/9409364\\September 20, 1994

\end{flushright}

\vskip1cm
\begin{center}
{\Large \bf Probing Factorization in Color-Suppressed $ B \rightarrow
\psi(2S) + K({K}^{*})$ Decay}\\[10mm]
A. N. Kamal and A. B. Santra\footnote{on leave of absence from
Nuclear Physics Division, Bhabha Atomic Research Centre, Bombay-400
085,  India.}\\[5mm]
{\em Theoretical Physics Institute and Department of Physics,\\
University of Alberta,
Edmonton, Alberta T6G 2J1, Canada.}
\end{center}

\begin{abstract}
In order to probe the factorization hypothesis in color-suppressed $B
\rightarrow \psi (2S) + K ({K}^{*})$ decays we have studied three
ratios:  $ R \equiv B (B \rightarrow \psi K) / B (B \rightarrow
\psi(2S) K)$, $ R' \equiv B (B \rightarrow \psi K^*) / B (B
\rightarrow \psi(2S) K^*)$ and $ P_L^\prime \equiv \Gamma_L (B
\rightarrow \psi(2S) K^*)/\Gamma (B \rightarrow \psi(2S) K^*)$ in
several scenarios for the $ q^2$-dependence of the form factors
involved. We find that measurements of R and $R' $, which exist, do
not yield scenario-independent tests of factorization. However, the
predicted range of $ P_L^\prime$ in all scenarios lies in a narrow
range, $ 0.50 {\
\lower-1.2pt\vbox{\hbox{\rlap{$<$}\lower5pt\vbox{\hbox{$\sim$}}}}\ }
P_L^\prime \leq 0.67$, and could test factorization.
\end{abstract}

PACS index: 13.25Hw, 14.40.Nd
\end{titlepage}

\section {Introduction}

The CP asymmetries in the color-suppressed processes, $B \rightarrow
\psi K $ and $ B \rightarrow \psi K^*$, could, in principle,
determine the angle $ \beta$ in the unitarity triangle\cite{R1}. For
this to be possible several conditions have to be met, one of which
being that these transitions be dominated by a single process - in
this case the factorized amplitude. These decays being
color-suppressed (amplitude proportional to the parameter $
a_2$\cite{R2}) have a smaller branching fraction than  the
color-favored modes such as $ B^0 \rightarrow D^- \pi^+$ and $
D^-\rho^+$\cite{R3}, whose decay amplitudes are proportional to the
parameter $ a_1$. Thus, a non-factorizable effect could, in
principle, contribute a larger fraction of the total amplitude to the
color-suppressed modes than to the color-favored ones.

Ref.[4] has shown that within the factorization scheme none of the
popular models of hadronic form factors explain the measured value of
the longitudinal polarization fraction$ {\Gamma_L \over \Gamma}\left(
B \rightarrow \psi K^* \right)$. Same conclusions have been reached
by the authors of ref.[5]. The implication is that if the experiments
are correct then either the factorization hypothesis breaks down or
the model-form factors and their $ q^2$ extrapolations are incorrect.

In order to test factorization in color suppressed decays of B into
charmonium one could use $ B \rightarrow \eta_c K\left( K^* \right)$
as has been done in ref.[6], or $ B \rightarrow \psi\left( 2S \right)
K \left( K^*\right)$ as we do here. To be useful, the value of the
testable quantity must span a reasonably narrow range and be as
independent of the assumptions on the form factors and their $q^2$
dependence as possible. As we will see in this paper the last
condition is hard to meet. Thus we deem our study to be more of a $
'$probe$'$  than a $ '$test$'$ of factorization scheme.

In order to study the predictions of the factorization scheme in $ B
\rightarrow \psi \left( 2S \right) + K \left( K^* \right)$ decays,
let us define three ratios,

\begin{eqnarray}
R&\equiv&{{B\left( B \rightarrow \psi K \right)} \over {B \left( B
\rightarrow \psi'  K\right)}}\\
R'&\equiv&{{B\left( B \rightarrow \psi {K}^{*} \right)} \over {B
\left( B \rightarrow \psi '  {K}^{*}\right)}}\\
P_L^\prime&\equiv&{\Gamma_L \over \Gamma}\left( B \rightarrow \psi'
K^* \right)
\end{eqnarray}
where $ \psi'$ is $ \psi \left( 2S \right)$.

In the factorization scheme the ratio R depends on $f_\psi$,
$f_{\psi'}$, both of which are known with reasonably small errors,
and how the form factor $F_1^{BK}$ extrapolates from $ q^2 =
m_\psi^2$ to $m_{\psi'}^2 $, and not its absolute value. We discuss
the details of the calculation in the next section.

Ratio $R'$, in the factorization scheme, depends, in addition to
$f_\psi$ and $f_{\psi'}$, on the following ratios\cite{R4},
\begin{eqnarray}
x \equiv {A_2^{BK^*} \left( m_\psi^2 \right)  \over A_1^{BK^*}
\left( m_\psi^2 \right) }\nonumber\\
y \equiv {V^{BK^*}\left( m_\psi^2\right) \over A_1^{BK^*}  \left(
m_\psi^2 \right) }
\end{eqnarray}
and
\begin{eqnarray}
x' \equiv {A_2^{BK^*} \left( m_{\psi'}^2 \right) \over A_1^{BK^*}
\left( m_{\psi'}^2 \right) } \nonumber\\
y' \equiv {V^{BK^*}\left( m_{\psi'}^2 \right) \over A_1^{BK^*}
\left( m_{\psi'}^2 \right) }
\end{eqnarray}

We determine the ranges of allowed values of $(x,y)$ and ($ x'$,$y'$)
as follows. First, using the weighted average of the longitudinal
polarization fraction, $ {\Gamma_L \over \Gamma}\left( B \rightarrow
\psi K^* \right)$ measured by CLEO\cite{R7} and CDF\cite{R8}, we
determine the allowed range of the parameter set $ (x,y)$ by the
method detailed in ref.[4]. To extrapolate $(x,y)$ to $ (x ',y ')$ we
use seven different assumed scenarios for the behavior of the form
factors $ A_1^{BK^*}$, $ A_2^{BK^*}$ and $V^{BK^*}$. This then
enables us to calculate the ratios $ R'$ and $ P_L^\prime$ of eq.(2)
and (3) in each scenario. Details follow below.

\section {Method and Results}

\underline{The Ratio R}:

We start with an estimate of the ratio R defined in eq.(1). In the
factorization scheme it is given by
\begin{eqnarray}
R_{th} = {f_\psi^2 \over f_{\psi'}^2} { \mid \vec{p}\mid^3 \over \mid
\vec{p'}\mid^3}  \Biggl | {F_1^{BK}\left( m_\psi^2 \right) \over
F_1^{BK}\left( m_{\psi'}^2 \right)}\Biggr |^2
\end{eqnarray}
where $\mid \vec{p}\mid$,  $\mid \vec{p'}\mid$ are the centre of mass
momenta in $ B \rightarrow \psi K$ and $ B \rightarrow \psi' K$
respectively. Using $ f_\psi = (384 \pm 14)$ MeV and $ f_{\psi'} =
(282 \pm 14)$ MeV\cite{R9} we obtain
\begin{eqnarray}
R_{th} = (4.13 \pm 0.51) \Biggl | {F_1^{BK}\left( m_\psi^2 \right)
\over F_1^{BK}\left( m_{\psi'}^2\right)}\Biggr |^2
\end{eqnarray}
Using now a monopole\cite{R2} or a dipole\cite{R9} extrapolation for
$F_1^{BK}$ with pole mass of 5.43 GeV, we get
\begin{eqnarray}
R_{th}&=&(2.64 \pm 0.33), monopole\nonumber\\
&=&(1.69 \pm 0.21), dipole
\end{eqnarray}
Experimentally this ratio is known, albeit with large errors, for
charged B decays\cite{R3}. From the listed branching fractions for $
B^+ \rightarrow \psi K^+$ and $ \psi' K^+$, we obtain
\begin{eqnarray}
R_{exp} = (1.48 \pm 0.69)
\end{eqnarray}
The actual error in $ R_{exp}$ is probably much less as some of the
systematic errors would cancel in the ratio. If we were to  reduce
the error by a third, a reasonable estimate of true error in eq.(9 )
is $ \pm 0.46$.

Using heavy quark symmetry it has been shown \cite{R9} that for heavy
to heavy flavor transitions such as $ B \rightarrow D$, the form
factor $F_1^{BD}$ ought to have a dipole structure though not
necessarily with identical pole masses. The authors of ref.[9] then
proceed to use a dipole form for $F_1^{BK}(q^2)$, despite the fact
that it is a heavy to light flavor transition, with some success.

A comparison of eq.(8) and eq.(9) shows that the theoretical value of
the ratio R calculated in the factorization scheme is in agreement
with its experimental determination provided that $F_1^{BK} $ has a
dipole behavior. A monopole form of $F_1^{BK} $ yields a value of R
too large by one theoretical and an experimental standard deviation.
If an independent case could be made (and we are not aware of one)
for a dipole form for $ F_1^{BK}(q^2)$ then our results would
constitute a test positive for the factorization hypothesis. On the
other hand, it must be said that a QCD sum rule-based
calculation\cite{R10} of  $ F_1^{B\pi}(q^2)$ is consistent with a
monopole form with a pole mass of (5.25 $\pm $ 0.1) GeV. The case of
$ F_1^{BK}(q^2)$ is anticipated to be similar but with a pole mass in
the region of 5.43 GeV. We are not aware of a QCD sum rule-based
calculation of $ F_1^{BK}$ and its $ q^2$ dependence.

 The discussion above does not establish nor contradict the
factorization hypothesis. The value of $ R_{th}$ depends on the
behavior of $ F_1^{BK}$ as a function of $ q^2$. Unfortunately,
unlike $ D \rightarrow K$ transitions, $ B \rightarrow K$ transition
involves a neutral current; consequently $ F_1^{BK}$ can not be
accessed through semileptonic B decays.

\underline{The ratio R$'$}:

In the factorization scheme, the ratio  $R'$ is given by,
\begin{eqnarray}
R'&=&{m_\psi^2 \over m_{\psi'}^2} {f_\psi^2 \over f_{\psi'}^2}
{\mid\vec{p}\mid \over \mid\vec{p'}\mid} \Biggl | {A_1^{BK^*}\left(
m_\psi^2 \right) \over A_1^{BK^*}\left( m_{\psi'}^2 \right)}\Biggr
|^2 f(x,y)\nonumber\\
&=&(1.844 \pm0.227) \Biggl | {A_1^{BK^*}\left( m_\psi^2 \right) \over
A_1^{BK^*}\left( m_{\psi'}^2 \right)}\Biggr |^2 f(x,y)
\end{eqnarray}
$ \mid\vec{p}\mid$ and $ \mid\vec{p'}\mid$ are the centre of mass
momenta in $ B \rightarrow \psi K^*$ and $ B \rightarrow \psi' K^*$
decays respectively, and
\begin{eqnarray}
f(x,y) = {\left( a - bx \right)^2 + 2\left(1 + c^2 y^2 \right) \over
\left( a' - b'x'\right)^2 + 2\left(1 + c'^2 y'^2 \right)}
\end{eqnarray}
$(x,y)$ and ($ x'$,$y'$) have been defined in (4) and (5) and, as
explained in the paragraph following (5),  ($ x'$,$y'$) are related
to $(x,y)$ by assumed scenarios of the $ q^2$-dependence of the form
factors.  $(a,b,c)$ and ($a' $,$ b'$,$ c'$) are defined as in
ref.[4],
\begin{eqnarray}
a&=&{m_B^2 - m_{K^*}^2 - m_\psi^2 \over 2m_{K^*}m_\psi} \nonumber\\
b&=&{2\mid\vec{p}\mid^2m_B^2 \over 2m_{K^*}m_\psi\left( m_B + m_{K^*}
\right)^2}\\
c&=&{2\mid\vec{p}\mid m_B \over \left( m_B + m_{K^*}
\right)^2}\nonumber
\end{eqnarray}
($a'$,$b'$,$c'$) are obtained from eq.(12) by replacing $m_\psi$ by
$ m_{\psi'}$ and $ \mid\vec{p}\mid$ by  $ \mid\vec{p'}\mid$.

Our procedure is to determine the allowed range of the parameters $x$
and $y$ from  $ {\Gamma_L\over \Gamma}\left( B \rightarrow \psi K^*
\right)$ data and then extrapolate them to  ($ x'$,$y'$) using
assumed $q^2$ dependence for the form factors.

For the longitudinal polarization fraction in $B \rightarrow \psi
K^*$ we use a weighted average of the following data:
\begin{eqnarray}
{\Gamma_L \over \Gamma}\left( B \rightarrow \psi K^*\right)&=&0.80
\pm 0.08 \pm0.05, CLEO\cite{R7} \nonumber\\
&=& 0.66 \pm 0.10_{-0.08}^{+0.10}, CDF\cite{R8}
\end{eqnarray}
with a weighted average,
\begin{eqnarray}
 {\Gamma_L \over \Gamma}\left( B \rightarrow \psi K^*\right) = 0.76
\pm 0.08,  \end{eqnarray}
where we have added the statistical and the systematic errors in
quadrature,

As shown in ref.[4], factorization assumption yields,
\begin{eqnarray}
 {\Gamma_L \over \Gamma}\left( B \rightarrow \psi K^*\right) =
{\left( a - bx \right)^2 \over \left( a - bx \right)^2 + 2\left( 1 +
c^2 y^2 \right)}

 \end{eqnarray}
The allowed region in $(x,y)$ plane determined from eqs.(14) and (15)
is shown in Fig.1, where we have also indicated the six points
representing the predictions of the six models (or schemes) referred
to in ref.[4]. Though we do not show this in Fig.1, CDF data by
themselves are in agreement with the predictions of models A and F of
Fig.1.

Having determined $x$ and $y$, we obtain the corresponding $ x'$ and
$ y'$ assuming the following seven different scenarios for the
extrapolation of the form factors:

\underline{Scenario 1}: $ A_1^{BK^*}$ and $ A_2^{BK^*}$ are flat;
$V^{BK^*}$ is a monopole with pole mass 5.43 GeV.

\underline{Scenario 2}: $ A_1^{BK^*}$ and $ A_2^{BK^*}$ are flat;
$V^{BK^*} $is a monopole with pole mass 6.6 GeV.

\underline{Scenario 3}: $ A_1^{BK^*}$ is flat, $V^{BK^*} $is a
monopole with pole mass 6.6 GeV, and  $ A_2^{BK^*}$ is rising slowly
and linearly in the $ q^2$ range: 0 $ \leq q^2 \leq m_{\psi'}^2$,
\begin{eqnarray}
{A_2^{BK^*}\left( q^2 \right) \over A_2^{BK^*}\left( 0 \right)} = 1 +
0.0222 {q}^{2}
\end{eqnarray}
with $ {q}^{2}$ in $ {GeV}^{2}$

This form is motivated by the calculation of $A_2^{B\rho}$ in
ref.[10] where this form factor rises almost linearly in the range of
$q^2$ we need. (The actual form we have used was non-linear; to
simulate the behavior of $A_2^{BK^*}\left( q^2 \right)/
A_2^{BK^*}\left( 0 \right) $ given in ref.[10], eq.(16) is a close
approximation in the range 0 $ \leq q^2 \leq m_{\psi'}^2$.)

\underline{Scenario 4}:  $ A_2^{BK^*}$ and  $V^{BK^*} $ are as in
scenario 3 and  $ A_1^{BK^*}$ is decreasing slowly and linearly in
the range 0 $ \leq q^2 \leq m_{\psi'}^2$,
\begin{eqnarray}
{A_1^{BK^*}\left( q^2 \right) \over A_1^{BK^*}\left( 0 \right)} = 1 -
0.313 q^2
\end{eqnarray}
with $ q^2$ in $ GeV^2$

This form was also motivated by the calculation of $A_2^{B\rho}$ in
ref.[10] where this form factor drops almost linearly in the range of
$ q^2$ that we need.

\underline{Scenario 5}:  $ A_1^{BK^*}$ and $ A_2^{BK^*}$ are
monopoles with pole masses 5.82 GeV, and $V^{BK^*}$ is a monopole
with pole mass 5.43 GeV. This is Model BSWI of ref.[4].

\underline{Scenario 6}:  $ A_1^{BK^*}$ is a monopole with pole mass
5.82 GeV,  $ A_2^{BK^*}$  is flat, and $V^{BK^*}$ is a monopole with
pole mass 5.43 GeV.

\underline{Scenario 7}:   $ A_1^{BK^*}$ is a monopole with pole mass
5.82 GeV;  $ A_2^{BK^*}$   and $V^{BK^*}$  are dipoles with pole
masses 5.82 GeV and 5.43 GeV respectively.  This is model BSWII of
ref.[4].

Once we have obtained the values of $ x'$ and $ y'$ in each of the
scenarios, we can  calculate $ R'$ and $ P_L^\prime$ of eq.(3) which
is given by
\begin{eqnarray}
P_L^\prime ={\left( a' - b'x' \right)^2 \over \left( a' - b'x'
\right)^2 + 2\left(1 + c'^2 y'^2 \right)}
\end{eqnarray}

In Table 1 we have shown our results for the predicted ranges of $
R'$ and $ P_L^\prime$ in each of the seven scenarios.

The stability of some numbers in Table 1 can readily be understood.
For example, it is evident from eq.(18) that the maximum value of  $
P_L^\prime$ occurs at $x'  = 0$, $y' = 0$, and hence is scenario
independent. The minimum occurs at (0,$y_{max}^\prime$). This is
scenario dependent, but, fortunately, this dependence is weak. Thus
independent of scenario for the $ q^2$-dependence of the form factor,
factorization scheme predicts that CLEO and CDF data for $ P_L$ imply
that $ P_L^\prime \leq 0.67$, and  $ P_L^\prime {\
\lower-1.2pt\vbox{\hbox{\rlap{$>$}\lower5pt\vbox{\hbox{$\sim$}}}}\ }
0.5$ though weakly scenario-dependent. It is this rather limited
range of $P_L^\prime $ that makes it a useful testing ground for the
factorization hypothesis.

As for the ratio $ R'$, it depends not only on the values of $
A_1^{BK^*}\left( m_\psi^2 \right)$ and $ A_1^{BK^*}\left( m_{\psi'}^2
\right)$ (see eq.(10)) but also on $f(x,y)$ (see eq.(11)). In
scenarios 1, 2 and 3, where $ A_1^{BK^*}$ is flat, $ R'$ depends only
on $f(x,y)$. The maximum of $f(x,y)$ occurs at $x = 0$ and $y = 0$
(this point maps to $ x'$ = 0, $ y'$ = 0) and, hence,
scenario-independent. The minimum of $f(x,y)$ occurs at ($
x_{max}$,0) (which maps to ($ x_{max}^\prime$,0)). Hence the minimum
is independent of $ V^{BK^*}$. Because of this, as scenarios 1 and 2
differ only in the behaviour of $ V^{BK^*}$, both the maximum and
minimum of $f(x,y)$ are identical and, consequently, also $ R'$.

{}From Table 1 it is evident that if  $ A_1^{BK^*}$ falls with $ q^2$,
$ R'$ rises and vice-versa.

{}From the branching ratios listed in ref.[3] for $ {B}^{0} \rightarrow
\psi K^*$ and $ \psi' K^*$, we obtain
\begin{eqnarray}
R_{exp}^\prime = 1.13 \pm 0.75
\end{eqnarray}

Again, due to partial cancellation of systematic errors the actual
error is probably lower.

Comparing the predicted values of $ R'$ in the seven different
scenarios shown in Table 1 with eq.(19), we can only conclude that if
factorization assumption were to hold, then the only scenarios that
are consistent with experiment are those in which $ A_1^{BK^*}$ rises
with $ q^2$; as to the behavior of  $ A_2^{BK^*}$ and $ V^{BK^*}$ we
learn little.

\begin{table}
\begin{center}
\caption{Range of $ R'$ and $ P_L^\prime$ in different scenarios}
\vspace{3mm}
\begin{tabular}{|c|c|c|}
\hline
     Scenario &$         R'$ &$       P_L^\prime$\\
\hline
1&2.23 $\leq R'  \leq$ 4.00&0.50 $\leq P_L^\prime \leq$ 0.67 \\
2&2.23 $\leq R'  \leq$ 4.00&0.53 $\leq P_L^\prime \leq$ 0.67 \\
3&2.29 $\leq R'  \leq$ 4.00&0.53 $\leq P_L^\prime \leq$ 0.67 \\
4&3.65 $\leq R'  \leq$ 5.92&0.48 $\leq P_L^\prime \leq$ 0.67 \\
5&1.56 $\leq R'  \leq$ 2.79&0.55 $\leq P_L^\prime \leq$ 0.67 \\
6&1.48 $\leq R'  \leq$ 2.79&0.55 $\leq P_L^\prime \leq$ 0.67 \\
7&1.66 $\leq R'  \leq$ 2.79&0.49 $\leq P_L^\prime \leq$ 0.67 \\
\hline
\end{tabular}
\end{center}
\end{table}

\section{Summary}
We have studied the color-suppressed $ B \rightarrow \psi(2S) +
K(K^*)$ in the factorization scheme. We calculated the three ratios
R, $ R'$ and $P_L^\prime $ defined in eqs.(1), (2) and (3)
respectively.
The ratio $ R_{th}$ agrees with $R_{exp}$ if $F_1^{BK}$ were to have
a dipole behavior in $q^2 $. On the other hand, if one could argue
(and we are not aware of such an argument) that $ F_1^{BK}$ has a
monopole behavior in the region 0 $ \leq q^2 \leq m_{\psi'}^2$, then
we would construe our result as signalling a break-down of the
factorization hypothesis. We do not claim to have tested
factorization.

In order to calculate $ R'$ and $P_L^\prime$ we first determined
$(x,y)$ from the measurement of $ P_L$, the longitudinal polarization
in $B \rightarrow \psi(2S) + K(K^*)$,  and then mapped them to ($
x'$,$ y'$) in seven assumed scenarios. We found that only in
scenarios where $ A_1^{BK^*}$ was assumed to rise at least as fast as
a monopole with pole mass 5.82 GeV did the calculated value of $ R'$
agree with experiment. This, however, does not constitute a test of
factorization.

The longitudinal polarization in $B \rightarrow \psi(2S) + K(K^*)$,
$ P_L^\prime$, however does have rather a limited range of values,
\begin{eqnarray}
0.5 {\
\lower-1.2pt\vbox{\hbox{\rlap{$<$}\lower5pt\vbox{\hbox{$\sim$}}}}\ }
P_L^\prime \leq 0.67,
\end{eqnarray}
the minimum showing only a weak dependence on the scenario. Thus a
measurement of $P_L^\prime$ could constitute a test of factorization.

Lastly, we wish to point out that whereas in ref.[4] where model
calculations of the form factors (both magnitude and $
q^2$-dependence) were confronted with data, here we needed to assume
only the $ q^2$-behavior of the form-factors in the region
$m_{\psi}^2 \leq q^2 \leq m_{\psi'}^2$ and not their absolute
normalization.
\vskip 0.5cm
ANK wishes to acknowledge a research grant fron the Natural Sciences
and Engineering Research Council of Canada which partially supported
this research.

\newpage

\noindent{\large \bf Figure Captions}

\begin{description}
\item[Fig.1]The allowed region (shaded) in the $(x,y)$ plane
determined by the polarization fraction $ {\Gamma_L \over
\Gamma}\left( B \rightarrow \psi K^* \right)$ (weighted average of
CLEO \cite{R7}  and  CDF \cite{R8}  data). The six points correspond
to the predictions of the models identified in ref.\cite{R4}. See
ref.\cite{R4} for further details.
\end{description}


\newpage

\begin{thebibliography}{99}
\bibitem{R1}
See, for example, P. Burchat in \underline{Particle Physics - The
Factory Era}, Eds. B.A.Campbell, A.N.Kamal, P.Kitching and
F.C.Khanna, World Scientific, Singapore, 1991.
\bibitem{R2}
M. Bauer, B. Stech and M. Wirbel, Z. Phys. 34, 103 (1987).
\bibitem{R3}
Particle Data Group, \underline{Review of Particle Properties}, Phys.
Rev. D50, 1173  (1994)
\bibitem{R4}
M. Gourdin, A. N. Kamal and X. Y. Pham, Paris Report No. PAR/ LPTHE/
94 -19  (to be published in Phys. Rev. Lett.)
\bibitem{R5}
R. Alkesan, A. Le Yaouance, L. Oliver, O. P\`{e}ne and J.-C. Raynal,
Orsay Report No. LPTHE-Orsay 94/15.
\bibitem{R6}
M.Gourdin, Y. Y. Keum and X. Y. Pham, Paris Report No. PAR/ LPTHE/
94-32, 1994.
\bibitem{R7}
M.S.Alam et al., CLEO collaboration, Phys. Rev. D50, 43(1994).
\bibitem{R8}
K. Ohl, Talk at Division of Particles and Fields (APS) Meeting,
Albuquerque, N.M., 2-6 August, 1994.
\bibitem{R9}
M. Neubert, V. Rieckert, B.Stech and Q. P. Xu in \underline{Heavy
Flavours}, Eds.. A. J. Buras and M. Lindner, World Scientific,
Singapore, 1992.
\bibitem{R10}
P. Ball, Phys. Rev. D48, 3190 (1993).
\end{thebibliography}
\end{document}